\documentclass[a4paper,10pt,showpacs,notitlepage,nofootinbib,floatfix,superscriptaddress,prd]{revtex4-1}
\pdfoutput=1
\usepackage{amsfonts}
\usepackage{amsmath}
\usepackage{graphicx}
\usepackage{dcolumn}
\usepackage{bm}
\usepackage{amssymb}%
\usepackage{slashed}
\usepackage{color}
\usepackage{youngtab}

\def\beq{\begin{equation}}
\def\eeq{\end{equation}}

\def\bea{\arraycolsep .1em \begin{eqnarray}}
\def\eea{\end{eqnarray}}

\def\s0#1#2{\mbox{\small{$ \frac{#1}{#2} $}}}
\def\0#1#2{\frac{#1}{#2}}

\def\grgl{\:\hbox to -0.2pt{\lower2.5pt\hbox{$\sim$}\hss}{\raise3pt\hbox{$>$}}\:}
\def\klgl{\:\hbox to -0.2pt{\lower2.5pt\hbox{$\sim$}\hss}{\raise3pt\hbox{$<$}}\:}

\def\lsim{\mathrel{\rlap{\lower4pt\hbox{\hskip1pt$\sim$}}
    \raise1pt\hbox{$<$}}}                
\def\gsim{\mathrel{\rlap{\lower4pt\hbox{\hskip1pt$\sim$}}
    \raise1pt\hbox{$>$}}}                

\newcommand{\drawsquare}[2]{\hbox{%
\rule{#2pt}{#1pt}\hskip-#2pt
\rule{#1pt}{#2pt}\hskip-#1pt
\rule[#1pt]{#1pt}{#2pt}}\rule[#1pt]{#2pt}{#2pt}\hskip-#2pt
\rule{#2pt}{#1pt}}
\newcommand{\Yfund}{\raisebox{-.5pt}{\drawsquare{6.5}{0.4}}}
\newcommand{\Ysymm}{\Yfund\hskip-0.4pt%
                    \Yfund}
\def\symm{\Ysymm}

\def\drawbox#1#2{\hrule height#2pt
        \hbox{\vrule width#2pt height#1pt \kern#1pt
              \vrule width#2pt}
              \hrule height#2pt}
\def\Fund#1#2{\vcenter{\vbox{\drawbox{#1}{#2}}}}
\def\Asym#1#2{\vcenter{\vbox{\drawbox{#1}{#2}
              \kern-#2pt 
              \drawbox{#1}{#2}}}}

\def\fund{\Fund{6.4}{0.3}}
\def\asymm{\Asym{6.4}{0.3}}

\begin{document}
\title{Viscous Conformal Gauge Theories}
\author{Arianna Toniato}
\affiliation{CP$^3$-Origins \& the Danish Institute for Advanced Study, Danish IAS, 
University of Southern Denmark, Campusvej 55, DK--5230 Odense, Denmark}
\author{Francesco Sannino}
\affiliation{CP$^3$-Origins \& the Danish Institute for Advanced Study, Danish IAS, 
University of Southern Denmark, Campusvej 55, DK--5230 Odense, Denmark}
\author{Dirk H.\ Rischke}
\affiliation{Institute for Theoretical Physics, Goethe University,
Max-von-Laue-Str.\ 1, D--60438 Frankfurt am Main, Germany }
\affiliation{Interdisciplinary Center for Theoretical Study and Department of
Modern Physics, University of Science and Technology of China, Hefei,
Anhui 230026, China}
\begin{abstract} 
We present the conformal behavior of the shear viscosity-to-entropy density ratio and the 
fermion-number diffusion coefficient 
within the perturbative regime of the conformal window for gauge-fermion theories.\\
[.1cm]
{\footnotesize  \it Preprint: CP$^3$-Origins-2016-052 DNRF90  }
\end{abstract}
\maketitle

\section{Introduction}

Gauge theories constitute the backbone of the standard model of particle interactions. 
Gauge theories exist in several different phases that are naturally classified according to the force 
measured between static sources. 
Knowledge of the phase diagram proves crucial when investigating extensions of the standard model both for 
particle physics and cosmology.  A special class of gauge theories are the ones that are fundamental according to 
Wilson \cite{Wilson:1971bg,Wilson:1971dh}, meaning that they possess a complete (in all couplings) ultraviolet 
(UV) fixed point either of non-interacting (asymptotically free 
\cite{Gross:1973id,Politzer:1973fx,Gross:1973ju,Cheng:1973nv,Callaway:1988ya}) or of interacting nature 
(asymptotically safe \cite{Weinberg:1980gg}).  Complete asymptotically safe quantum field theories were discovered only 
very recently \cite{Litim:2014uca,Litim:2015iea}, widening the horizon of fundamental 
theories\footnote{A crucial property was unveiled in Ref.\ \cite{Litim:2014uca}, i.e., the Yukawa interactions, 
mediated by the scalars, compensate for 
the loss of asymptotic freedom due to the large number of gauged fermion flavors and 
therefore cure the subsequent growth of the gauge coupling. 
The further interplay of the gauge, Yukawa, and scalar interactions ensures that all 
couplings reach a stable interacting UV fixed point, allowing for a 
{\it complete asymptotic safety} scenario in all couplings \cite{Litim:2014uca}. 
This is different from the {\it complete asymptotic freedom} scenario 
\cite{Gross:1973ju,Cheng:1973nv,Callaway:1988ya} where all couplings 
vanish in the UV. } that can be used for novel phenomenological applications \cite{Sannino:2014lxa} beyond the 
traditional asymptotically free paradigm \cite{Gross:1973id,Politzer:1973fx}. The thermal 
properties of completely asymptotically safe field theories were elucidated in Ref.\ \cite{Rischke:2015mea}. 

Here we focus our attention on asymptotically free gauge theories featuring gauge and fermion degrees of freedom that 
develop an infrared (IR) interacting fixed point.  We henceforth push forward our program to systematically understand, 
in a rigorous manner, the dynamics of these theories at zero 
\cite{Sannino:2010ca,Pica:2010mt,Pica:2010xq,Ryttov:2010iz,Ryttov:2016ner,Pica:2016rmv} and non-zero matter density 
\cite{Sondergaard:2011ps,Mojaza:2010cm}, by analysing their conformal viscous behavior as a function of the number 
of flavors\footnote{Systematic analytic studies of the conformal window of 
non-supersymmetric field theories beyond perturbation theory
re-started in Refs.\ \cite{Sannino:2004qp,Dietrich:2006cm}. Here the reader will also 
find a complete list of earlier references.}. Because of the perturbative nature of the theories investigated here, 
along the full energy range, our investigation of their viscous properties  is also much better controlled 
than for QCD-like theories. This is so because at very high energies the theory is non-interacting and at very low
energies the theory reaches an IR perturbative fixed point. Furthermore, the value of the gauge coupling at
the IR fixed point can be 
made arbitrarily small by changing the number of flavors and colors of the theory. 
This allows us to consistently truncate the perturbative expansion within the range of convergence of the 
theory. 

We henceforth determine the conformal behavior, as a function of the number of flavors, for the shear viscosity-to-entropy 
density ratio and the fermion-number diffusion coefficient. By adapting the results of Ref.\ \cite{Arnold:2000dr} we learn that, 
as we decrease the number of flavors below the loss of asymptotic freedom, their IR fixed point values decrease. 
Furthermore, for a given number of flavors within the perturbative conformal window both coefficients decrease with 
decreasing temperature (once we multiply the diffusion coefficient by the temperature) from their infinite value 
in the deep 
UV down to the value at the IR fixed point. We represent the results for three colors as a function of the number of 
flavors, but to the order investigated here the results are similar for any other fermion representation.

We organise this paper as follows. In Sec.\ \ref{sec:g2} we shortly review the theory, introduce the notation, and 
provide the salient zero and non-zero temperature properties.  This is followed by the determination of the transport 
coefficients in Sec.\ \ref{sectionIII}. Here we will comment on our findings and finally conclude in Sec.\ \ref{conclusions}.

\section{Review of the Hot Conformal Free Energy Density @ $\mathbf{\cal O}(g^2)$ and Entropy Density}\label{sec:g2} 

Our starting point is a generic asymptotically free gauge theory with $N_f$ Dirac flavors transforming according to 
the representation $r$ of the underlying gauge group.  
 
The relevant group-normalization factors are:
\begin{equation}
\text{Tr}[ T^a_r T^b_r ]  = T[r] \delta^{ab}\;, \qquad 
T^a_r T^a_r = C_2[r] \mathbf{1}\; ,
\end{equation}
where $T^a_r$ is the $a$-th group generator in the representation $r$ and  $a=1,\dots, d[G]$. We denote with $d[r]$ 
the dimension of the representation, and with $G$ the adjoint representation. The quantities $T[r]$ and $C_2[r]$ are 
related via the identity $C_2[r] d[r] = T[r] d[G]$. We summarise useful group theory factors in Table~\ref{factors}. 
    \begin{table}
\begin{center}
    \begin{tabular}{c||ccc }
    $r$ & $ \quad T(r) $ & $\quad C_2(r) $ & $\quad
d(r) $  \\
    \hline \hline
    $ \fund $ & $\quad \frac{1}{2}$ & $\quad\frac{N^2-1}{2N}$ &\quad
     $N$  \\
        $\text{$G$}$ &\quad $N$ &\quad $N$ &\quad
$N^2-1$  \\
        $\symm$ & $\quad\frac{N+2}{2}$ &
$\quad\frac{(N-1)(N+2)}{N}$
    &\quad$\frac{N(N+1)}{2}$    \\
        $\asymm$ & $\quad\frac{N-2}{2}$ &
    $\quad\frac{(N+1)(N-2)}{N}$ & $\quad\frac{N(N-1)}{2}$
    \end{tabular}
    \end{center}
\caption{Relevant group factors for the representations used
throughout this paper. However, a complete list of all the group
factors for any representation and the way to compute them is
available in Table II and the appendix of Ref.\
\cite{Dietrich:2006cm}.}\label{factors}
    \end{table}

The $\beta$ function up to four-loop order,
 \begin{align}
 \beta (g) = - \frac{\beta_0}{(4\pi)^2} g^3 - \frac{\beta_1}{(4\pi)^4} g^5
 -\frac{\beta_2}{(4\pi)^6} g^7 - \frac{\beta_3}{(4\pi)^8} g^9 + \mathcal{O}(g^{11})\; ,
 \end{align}
was computed in Ref.\ \cite{vanRitbergen:1997va}. As is the case for the free energy, the four-loop
$\beta$ function  is also 
computed in the $\overline{\text{MS}}$ scheme, thus
no ambiguities in the scheme dependence of the expressions arise. 
 Only $\beta_0$ and $\beta_1$ are scheme-independent and read: 
 \begin{align}
\beta_0 &= \frac{11}{3} C_2[G] - \frac{4}{3} T[r] N_f\; , \\
\beta_1 &= \frac{34}{3} C_2^2[G] - \left ( \frac{20}{3} C_2[G] + 4 C_2[r] \right ) T[r] N_f\; .
\end{align}
Asymptotic freedom is lost when the lowest-order coefficient, $\beta_0$,
changes sign. This occurs for
\begin{align}
N_f^{\rm AF} = \frac{11}{4} \frac{C_2[G]}{T[r]}\; .
\end{align}
For a given fermion representation, the second coefficient, $\beta_1$, is negative below and near this
critical number of flavors and an IR-stable fixed point develops, which is known as the Banks-Zaks 
fixed point \cite{Banks:1981nn}.
Such a theory displays large-distance conformality.  
The value of the coupling at the IR fixed point, $g_*$, is such that $\beta(g_*) = 0$, and it is given at 
next-to-leading order by:
\begin{equation}
g_*^2 = - (4 \pi)^2 \frac{\beta_0}{\beta_1} \;.
\end{equation} 
The IR fixed point disappears, at two-loop level, when $\beta_1$ changes sign. This occurs for: 
\begin{align}
\label{eq:NfIII}
N_f^{\rm Lost} = \frac{17 C_2[G]}{10 C_2[G] + 6 C_2[r]} \frac{C_2[G]}{T[r]}\; .
\end{align}

The free energy density is known up to the order $g^6 \log(1/g)$  \cite{Kajantie:2002wa} but for this exploratory 
study it is sufficient to stop at order $g^2$, where it reads:
\begin{align}
\frac{f}{\pi^2 T^4 } = - \frac{d[G]}{9} \left[ \frac{1}{5} + \frac{7}{20}\frac{d[r]}{d[G]}N_f 
-   \left ( C_2[G] + \frac{5}{2}T[r] N_f \right ) \frac{g^2(T)}{(4\pi)^2} \right ]\; ,
\end{align}
where $T$ is the temperature of the theory and we traded the renormalization scale by $T$. 
In the deep UV, i.e., at temperatures sufficiently high that the physics is dominated by the asymptotically 
free fixed point, the coupling vanishes logarithmically and the UV free energy density is the one of a free gas of gluons 
and fermions: 
\begin{align}
\frac{f^{UV}}{\pi^2 T^4 } = - \frac{d[G]}{9} \left[ \frac{1}{5} + \frac{7}{20}\frac{d[r]}{d[G]}N_f  \right]\; .
\end{align}
This is the trivial {\it conformal} limit while the interacting {\it conformal} free energy density 
in the deep IR is obtained by replacing 
the coupling constant with the Banks-Zaks fixed point value $g_*$ \cite{Mojaza:2010cm}:
\begin{align*}
\frac{f^{IR}}{\pi^2 T^4 } = \frac{f_*}{\pi^2 T^4 } =
- \frac{d[G]}{9} \left [ \frac{1}{5} + \frac{7}{20}\frac{d[r]}{d[G]}N_f 
+ \frac{\left ( C_2[G] + \frac{5}{2}T[r] N_f \right ) \left ( 11 C_2[G] - 4 T[r] N_f \right)}{
34 C_2^2[G] - ( 20 C_2[G] + 12 C_2[r] ) T[r] N_f} \right ]\; ,
  \end{align*}
We observe immediately that due to the {\it conformal} large-distance nature of our theories the dependence of
the free energy
density 
on the energy scale is only via the temperature, which factors out leaving behind  a numerical factor containing 
information on the specific theory studied. 


The entropy density $s$ can be determined via its relation with the free energy density: 
\begin{equation}
\frac{s}{4\pi^2 T^3} = - \frac{1}{4\pi^2 T^3}\frac{df}{ dT}  =    \hat{f} +  \frac{ \beta(g)}{4}\frac{\partial \hat{f}}{\partial g} 
 \; , 
\end{equation}
with $f = -{\hat{f}(g(T))} \pi^2 T^4$. At fixed points, where the $\beta$ function vanishes, 
\begin{equation}
\frac{s^{FP}}{4\pi^2 T^3} =  - \frac{f^{FP}}{\pi^2 T^4} \; .
\end{equation}
 
Having at our disposal the precise expressions of both the entropy and free energy density we can now move to the 
transport coefficients that encode further important dynamical properties of the theory.

\section{Flavor and Temperature Dependence of The Conformal Shear 
Viscosity and Fermion-Number Diffusion Coefficients}
\label{sectionIII}

We are now ready to unveil the dependence on the number of flavors for relevant transport coefficients such as the 
shear viscosity and fermion-number diffusion coefficient for several gauge theories at perturbatively 
trustable interacting fixed points. We will also analyse the temperature dependence of the mentioned transport 
coefficients, once the number of flavors and colors are fixed to some value in the perturbative conformal window.
 
In order to determine the transport coefficients, the authors of Refs.\ \cite{Arnold:2000dr,Arnold:2003zc} used 
kinetic theory in which coupled Boltzmann equations 
describe the evolution of the phase-space density of distinct particle species.  The transport coefficients can be read 
off from the stress-energy tensor of the theory, which in turn is determined once the phase-space densities of all the particle 
species are known. In Refs.\ \cite{Arnold:2000dr,Arnold:2003zc} analytic expressions for the transport 
coefficients are given, which approximately reproduce the numerical results. The result for the shear viscosity, 
in the next-to-leading-log approximation, is:  
\begin{equation}   
\eta \simeq 270 \, d [G] \zeta(5)^2 \biggl( \frac{2}{\pi} \biggr)^5 (v^T c^{-1} v) \frac{T^3}{g(T)^4 \ln( A\, T/ m_D)}\;,
\label{eta}
\end{equation}
where  
\[
c = (d [G] C_2 [G] + N_f d [r] C_2 [r])
\begin{pmatrix}
d [G] C_2 [G] & 0 \\
0 & \frac{7}{4} N_f d [r] C_2 [r]
\end{pmatrix}
+ \frac{9 \pi^2}{128} N_f d [r] C_2^2 [r] d [G]
\begin{pmatrix}
1 & -1 \\
-1 & 1
\end{pmatrix} \;,
\]

\[
v = 
\begin{pmatrix}
d [G] \\
\frac{15}{8} N_f d [r]
\end{pmatrix}\; ,
\]
 
 \begin{equation}
	\quad m_D^2 =  \frac{1}{3} \left(C_2[G] + N_f C_2[r]\frac{d[r]}{d[G]}  \right)g^2 T^2 \; ,
	\label{log} 
\end{equation}
with $m_D$ the Debye mass and $A$ a numerical coefficient that has a mild dependence on the number of 
flavors and colors.  The numerical values of $A$ relevant for the cases studied in this paper 
are reported in Table~\ref{AB}.

    \begin{table}
\begin{center}
    \begin{tabular}{c||ccc }
    $N_f$ & $ \quad A$ & $\quad B $ &   \\
    \hline \hline
    $ 6 $ & \quad 2.918 & \quad 3.064   \\
        $14$ &\quad 2.878 &\quad 3.135  \\
        $15$ & \quad 2.873 & \quad 3.172 \\
        $16$ & \quad 2.869  & \quad 3.176 \\
    $16.25$ & \quad 2.867 & \quad 3.177
    \end{tabular}
    \end{center}
\caption{Values of the coefficients $A$ and $B$ \cite{privcommGDM}
appearing in the next-to-leading-log expressions of the shear viscosity and the fermion-number diffusion coefficient, 
for $N = 3$ and different values of $N_f$.}\label{AB}
    \end{table}

Because of the overall $T^3$ dependence of the shear viscosity it is convenient to normalise it to the entropy density. 
The so constructed ratio reads at a generic fixed point
\begin{equation}
\frac{\eta^{FP}}{s^{FP} }
 = \frac{{\cal A}(N_ f, N) }{g_*^4 \ln[{\cal B}(N_f, N)g_*^{-1}]}  \; , 
\label{etaoversFP}
\end{equation}
with $ {\cal A}(N_f, N) $  and $ {\cal B}(N_f, N) $ calculable definite positive and smooth functions of the number of 
colors and flavors, with $g_* = g_* (N_f, N)$ the value of the coupling at the fixed point.  

As expected at non-interacting fixed points,  such as the UV fixed point, the ratio diverges. On the  other hand at the 
interacting IR fixed point the ratio approaches a finite value controlled by a small non-vanishing $\delta = N_f^{AF} - N_f $.

\begin{figure}[h!]
\includegraphics[width=0.46\textwidth]{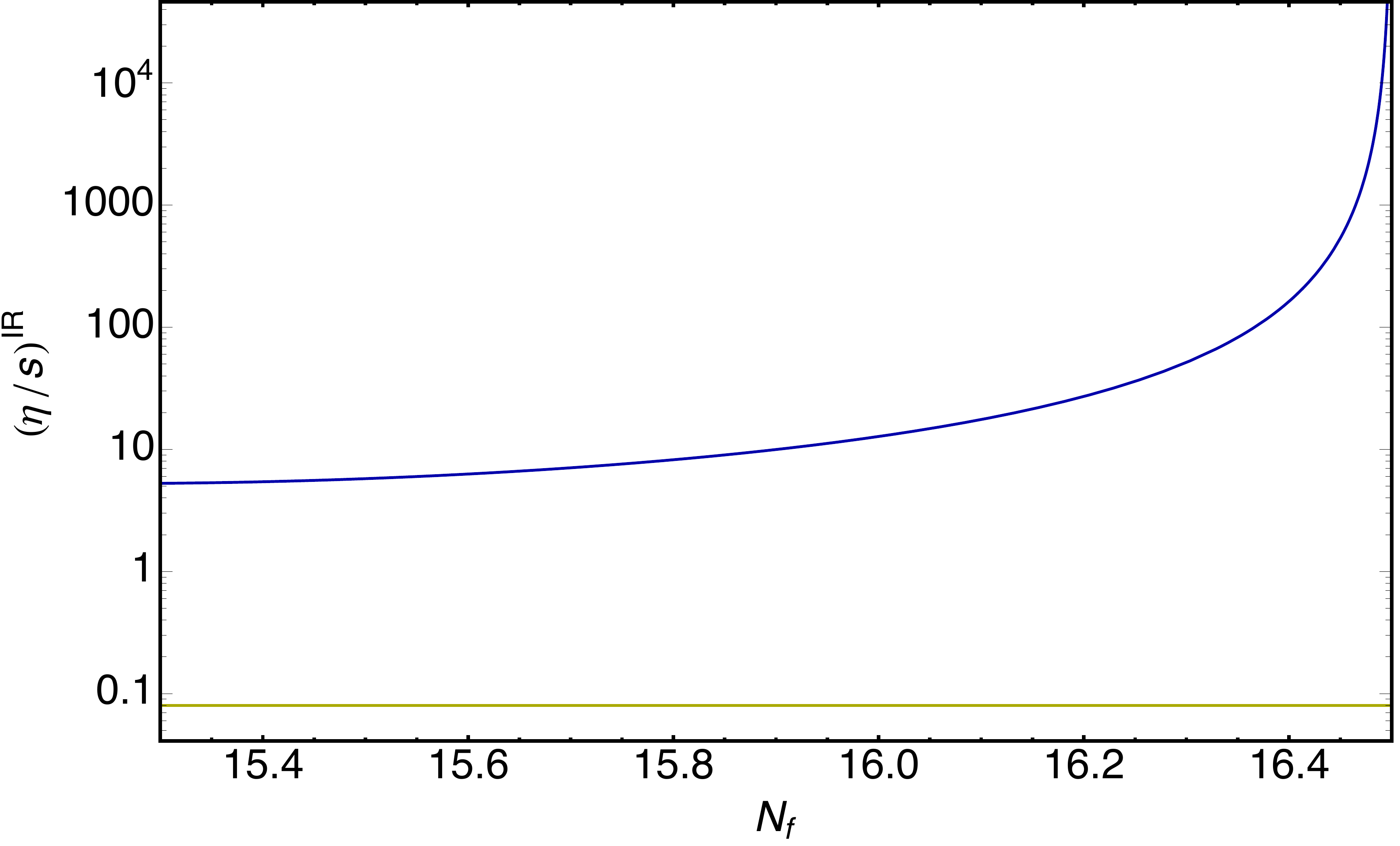}~~~
\includegraphics[width=0.46\textwidth]{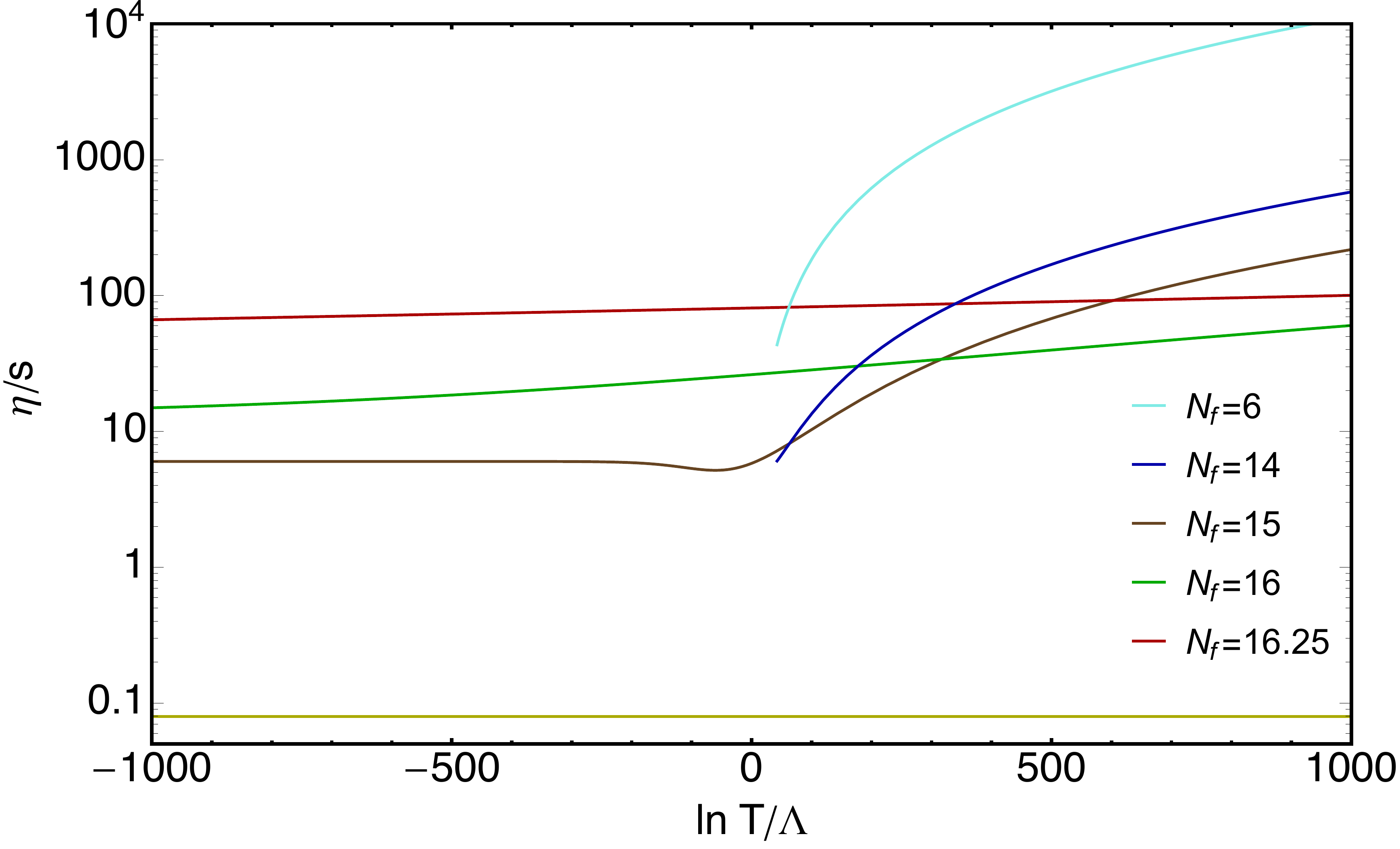}
\caption{Left Panel: ${\eta}/{s}$ evaluated at the IR fixed point, as a function of the number of flavors, for fermions in 
the fundamental representation with $N = 3$ colors. Right Panel:  ${\eta}/{s}$ as function of the temperature over the 
RG scale $\Lambda$  for different values of $N_f$ in the conformal window and one 
outside corresponding to $N_f = 6$, for $N=3$ colors. Although $N_f = 14$ still displays a potential IR fixed point 
the IR dynamics  of ${\eta}/{s}$ cannot be accessed perturbatively.  
The horizontal line at the bottom is the conjectured 
AdS/CFT bound.} \label{etas_Nf}
\end{figure}

In the left panel of Fig.~\ref{etas_Nf} we plot $(\eta/s)^{IR}$ as function of the number of flavors, for 
fermions in the fundamental representation with $N = 3$. When decreasing the number of flavors below the 
asymptotically free boundary, where the shear viscosity diverges, we observe a dramatic decrease while still remaining  
much above the bound $\eta/s \geq 1/(4\pi)$ conjectured by AdS/CFT \cite{Kovtun:2004de}. It is natural to expect that, as 
we further decrease the number of flavors, the IR ratio further decreases to reach a minimum value at the lower boundary 
of the conformal window. Below this critical number of flavors we expect the onset of chiral symmetry breaking and the 
theory in the deep IR becomes a theory of non-interacting pions with again a divergent value of this quantity.

In the right panel of Fig.~\ref{etas_Nf} we present the temperature dependence of the shear viscosity over the entropy 
density for 
several values of $N_f$. The quantity $\eta/s$ depends on the temperature over a reference scale $\Lambda$ via the 
gauge coupling. The reference energy scale is chosen to be the one for which the $\beta$ function displays a minimum 
occurring between the trivial UV and interacting IR fixed points. The energy scale $\Lambda$ is therefore defined by:
\begin{equation}
g^2(T = \Lambda)=\frac{3}{5}g^2_* \, .
\label{Lambda}
\end{equation} 
For $N_f = 6$, for which the theory does not display an IR perturbative fixed point, $\Lambda$ is taken to be the scale at 
which the one-loop gauge coupling diverges as function of the temperature.
The ratio $\eta/s$ decreases as we decrease the temperature for different values of 
the number of flavors within the conformal window.  However for $N_f=15$ we observe that a minimum develops 
around $T = \Lambda$.  This happens  because for this value of $N_f$ there is a temperature for which 
$4 \ln \left(\frac{AT}{m_D} \right)  = 1 $, which corresponds to a mininum for the 
$g^{-4} \ln \left(\frac{AT}{m_D} \right)  ^{-1} $ function. 

We now move our attention to another relevant transport quantity, the fermion-number diffusion coefficient. 
 The diffusion coefficient for the net number density of the fermion flavor $a$ is 
given in Ref.~\cite{Arnold:2000dr} and reads, at the next-to-leading-log level: 
\begin{equation}
D_a = \frac{6^5 \zeta(3)^2}{\pi^3 C_2[r_a]} \biggl[ \sum_b^{f \bar{f} h} T[r_b] \lambda_b + \frac{3 \pi^2}{8} 
C_2[r_a] \biggr]^{-1}  \frac{T^{-1}}{g^4 \ln(B\, T/ m_D)}\;, 
\label{diff_coeff}
\end{equation}
where the sum extends over all particle species $b$ that the fermion species $a$ can scatter 
with in the process $ab \to ab$, 
mediated by a gauge boson. Particles and antiparticles are counted separately, and the same goes for the helicity states: 
this means that we have to count a factor of four for every Dirac fermion, and a factor of two for gauge bosons. 
Furthermore, $\lambda_b = 1$ if the particle $b$ is a fermion, and $\lambda_b = 2$ if it is a boson. $B$ is a numerical coefficient, whose values relevant for the cases studied in this paper are reported in Table~\ref{AB}.

We can specialize Eq.\ \eqref{diff_coeff} to our theory with $SU(N)$ gauge symmetry and $N_f$ fermions, all in the 
same representation $r$. We obtain:
\begin{equation}
D = \frac{6^5 \zeta(3)^2}{\pi^3 C_2[r]} \biggl[ 4 N_f T[r] + 4 N + \frac{3 \pi^2}{8} C_2[r] \biggr]^{-1} 
 \frac{T^{-1}}{g^4\ln(B \, T/ m_D)} \;.
\end{equation}

\begin{figure}[h!]
\includegraphics[width=0.46\textwidth]{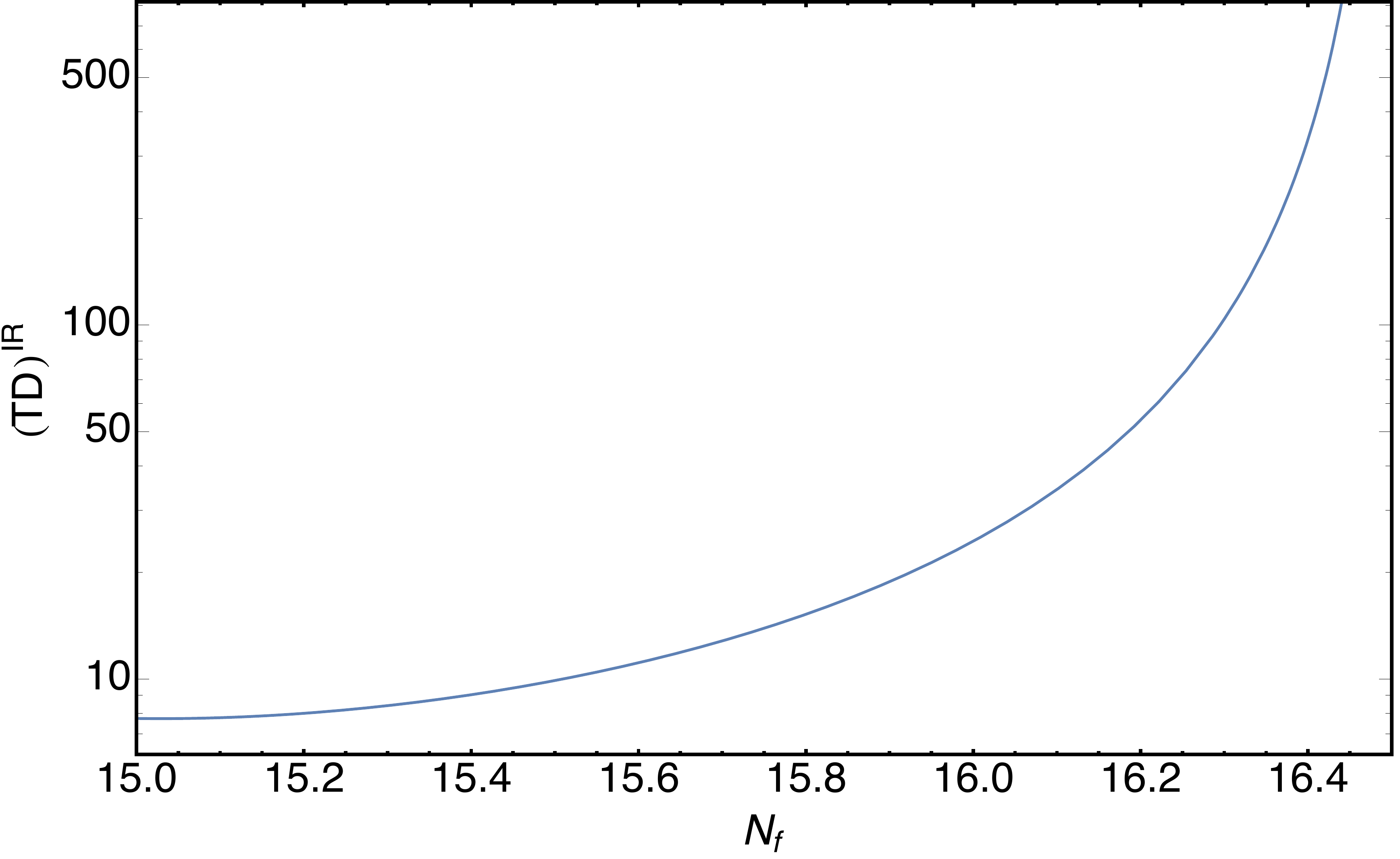}~~~
\includegraphics[width=0.46\textwidth]{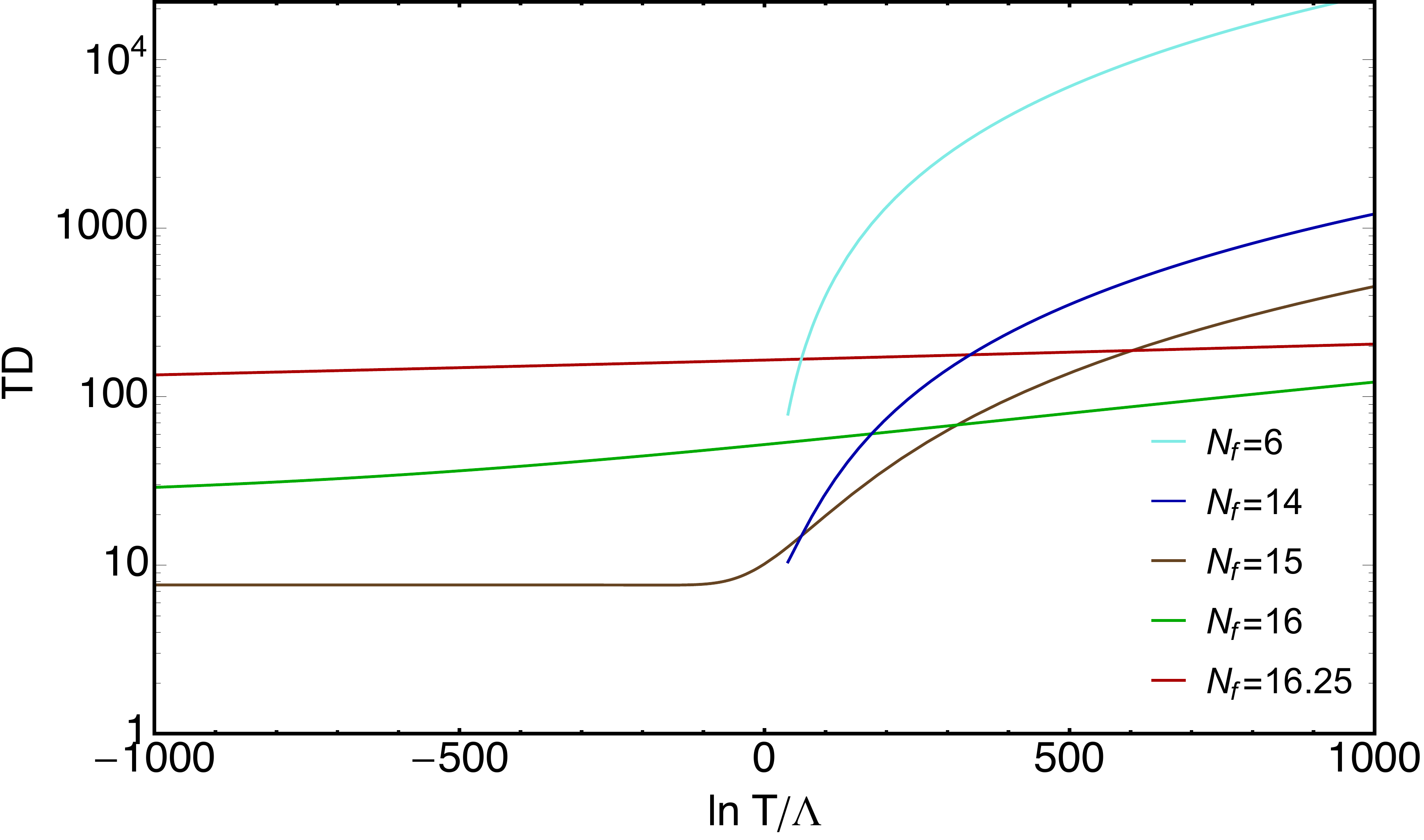}
\caption{Left Panel: $(T D)^{IR}$ as a function of the number of flavors, for the case of fermions in the fundamental 
representation and $N=3$. Right Panel: $T D$ as a function of the temperature, for different values of $N_f$ 
and $N=3$. }
\label{D_Nf}
\end{figure}

At very low energies, where the coupling is frozen at the fixed-point value $g_*$, the dimensionless quantity 
$(T D)^{IR}$ can be plotted as a function of the number of flavors. This is represented in the left panel of Fig.\
\ref{D_Nf}, for the case of fermions in the fundamental representation and $N=3$.  
As for the case of the shear viscosity-to-entropy density ratio, we observe that $(TD)^{IR}$ diverges as 
$g_*$ approaches the origin when increasing the number of flavors towards the asymptotic freedom boundary. 
As for the shear viscosity-to-entropy density ratio, in the right panel of Fig.\ \ref{D_Nf}
we also plot $T D$ as a function of temperature for different values 
of the number of flavors in the conformal window and for $N_f=6$.

One last comment has to be made about the applicability of the next-to-leading-log approximation for the 
transport coefficients in the conformal window. The presence of a perturbative IR fixed point allowed us to apply the 
next-to-leading-log results in the whole energy range, from the UV, where the theory is asymptotically free, down to the IR.
However, particular care has to be taken to decide whether the values obtained in the deep IR can be trusted. 
We chose to illustrate the results for the case of three colors and for different values of the number of flavors 
within the perturbative conformal window. $N_f=15$ is the last value at which we could observe the expected behavior of 
the transport coefficients as a function of the temperature, i.e., to run from a divergent value in the UV down to a constant 
finite value in the IR. For $N_f=14$ the next-to-leading-log expression for the transport coefficients does not stabilise at a 
finite value in the IR, but instead diverges at low energies, showing that the next-to-leading-log approximation cannot be 
trusted any longer. In fact, following Ref.\ \cite{Arnold:2003zc} one can argue that the next-to-leading-log result is very 
close to the full leading-order result (and therefore trustable) as long as $m_D/T \leq 1$.  This requirement is satisfied in our 
analysis provided $N_f$ is larger than $16.25$, de facto further limiting the window of applicability of the perturbative 
analysis.  The values of $m_D/T$ at the IR fixed point for $N=3$ and the values of $N_f$ within the conformal window
 that have been considered in this paper are reported in Table~\ref{mDTconstraint}.

   \begin{table}
\begin{center}
        \begin{tabular}{c||c }
    $N_f$ & $ \quad (m_D/T)^{IR}$ \\
    \hline \hline
        $14$ &$\quad 3.41$   \\
        $15$ & $\quad 2.51$ \\
        $16$ & $\quad 1.38$  \\
    $16.25$ & $\quad 0.97$ \\
    \end{tabular}
     \end{center}
\caption{Values of the ratio $m_D/T$ evaluated at the IR fixed point for $N=3$ and different values of $N_f$ in the 
conformal window. It can be observed that the constraint $m_D/T \leq 1$ is respected only in the case $N_f=16.25$.}
\label{mDTconstraint}
    \end{table}

\section{Conclusions}
\label{conclusions}
 
We determined the shear viscosity-to-entropy density ratio and the fermion-number diffusion coefficient within the 
perturbative regime of the conformal window for gauge-fermion theories.  
Our formalism is valid for any fermionic matter representation, while the physical results, which are 
expected to hold 
generically, were elucidated via a three-color  gauge theory as functions of the number of flavors in the fundamental 
representation.
We observed that when the number of 
flavors decreases from the value at the loss of asymptotic freedom both the shear viscosity-to-entropy density ratio and the 
fermion-number diffusion coefficient measured at the IR fixed point dramatically decrease. Furthermore, 
for a given number of flavors within the perturbative conformal window both coefficients decrease 
(albeit not  monotonically for $N_f=15$) with the 
temperature  from their divergent value in the 
UV down to the value at the IR fixed point. 
More specifically we discovered that  down to 15 flavors the next-to-leading-log results exhibit the expected 
behavior of stabilising at a constant finite value in the IR. For $N_f=14$ the next-to-leading-log results diverge at low energy, 
showing that the next-to-leading-log approximation cannot be trusted even qualitatively. In fact, following Refs.\ 
\cite{Arnold:2000dr,Arnold:2003zc} one can consider a more restrictive constraint for the next-to-leading-log 
approximation to be quantitatively accurate. The latter requires  $m_D/T \leq 1$ which, in our analysis, 
is valid for $N_f$ larger than $16.25$.

The ratio $\eta/s$ at the IR fixed point drops significantly when going from 16.25 to 
15 flavors showing that a modest change in the number of flavors dramatically affects the dynamics of the 
theory encoded in the transport 
coefficients. Higher-order corrections are needed to reach lower values of $N_f$ within the conformal window for the 
transport coefficients. In contrast, at zero temperature one observes that perturbation theory allows to go quite low in the 
number of flavors within the conformal window~\cite{Pica:2010mt,Pica:2010xq,Ryttov:2010iz,Ryttov:2016ner}. 
Although unproven it is reasonable to expect that the minimum of $\eta/s$ as function of temperature in QCD lies 
below the lowest value of $\eta/s$ obtained at the bottom of the conformal window, and therefore lower than 
the one obtained near 15 flavors. 

To conclude, the salient results of our analysis are: 
\begin{itemize} 
\item We provided theoretically relevant examples in which the perturbative estimate of the transport 
coefficients can be used along the entire RG flow from the UV to the IR without loosing their validity. 
\item We determined the range of applicability of those results within the conformal window of QCD and QCD-like theories.  
\end{itemize}
Our computations delineate and extend the range of applicability of the perturbative transport coefficients to the relevant 
subject of the conformal window in QCD and QCD-like theories. The work serves as stepping stone for future studies in 
this direction.

\section*{Acknowledgments}
We thank Guy D.\ Moore for providing the numerical values of the constants $A$ and $B$ given in Table II.
 The work of F.S.\ and A.T. \ is partially supported by the Danish National Research Foundation under the grant DNRF:90. 
D.H.R.\ is supported in part
by the High-End Visiting Expert project GDW20167100136 of the State Administration
of Foreign Experts Affairs (SAFEA) of China. 
\appendix

\end{document}